\documentclass[runningheads]{llncs}
\usepackage{times}
\usepackage{soul}
\usepackage{url}
\usepackage[hidelinks]{hyperref}
\usepackage[utf8]{inputenc}
\usepackage[small]{caption}
\usepackage{graphicx}
\usepackage{amsmath}
\usepackage{booktabs}
\usepackage{enumitem}
\usepackage{graphicx}
\usepackage{comment}
\usepackage{amssymb}
\usepackage[ruled,vlined,linesnumbered]{algorithm2e}
\begin{document}

\title{Multi-Agent Path Finding with Capacity Constraints}

\author{Pavel Surynek\inst{1}\orcidID{0000-0001-7200-0542} \and
T. K. Satish Kumar\inst{2} \and
Sven Koenig\inst{3}}
%

%\author{Pavel Surynek\inst{1}\orcidID{0000-0001-7200-0542}. T. K. Satish Kumar, Sven Koenig}
%\author{Paper number 19}
%
\authorrunning{P. Surynek et al.}


\institute{Faculty of Information Technology, Czech Technical University in Prague\\Th\'{a}kurova 9, 160 00 Praha 6, Czechia\\
\email{pavel.surynek@fit.cvut.cz}\\
\url{http://users.fit.cvut.cz/surynek} \and
University of Southern California, Henry Salvatori Computer Science Center\\941 Bloom Walk, Los Angeles, USA\\
\email{skoenig@usc.edu} \and
University of Southern California, Information Sciences Institute\\4676 Admiralty Way, Marina del Rey, USA\\
\email{tkskwork@gmail.com}
}

%\institute{
%Faculty of Information Technology\\
%Czech Technical University in Prague\\
%Th\'{a}kurova 9, 160 00 Praha 6, Czech Republic\\
%\email{pavel.surynek@fit.cvut.cz}
%}
%
\maketitle              % typeset the header of the contribution

%\author{Pavel Surynek}
%\orcid{0000-0001-7200-0542}
%\affiliation{
%  \institution{Faculty of Information Technology\\Czech Technical University}
%  \streetaddress{Th\'{a}kurova 9}
%  \city{Praha} 
%  \state{Czechia} 
%  \postcode{160 00}
%}
%\email{pavel.surynek@fit.cvut.cz}

\begin{abstract}
In multi-agent path finding (MAPF) the task is to navigate agents from their starting positions to given individual goals. The problem takes place in an undirected graph whose vertices represent positions and edges define the topology. Agents can move to neighbor vertices across edges. In the standard MAPF, space occupation by agents is modeled by a capacity constraint that permits at most one agent per vertex. We suggest an extension of MAPF in this paper that permits more than one agent per vertex. Propositional satisfiability (SAT) models for these extensions of MAPF are studied. We focus on modeling capacity constraints in SAT-based formulations of MAPF and evaluation of performance of these models. We extend two existing SAT-based formulations with vertex capacity constraints: MDD-SAT and SMT-CBS where the former is an approach that builds the model in an eager way while the latter relies on lazy construction of the model.

\end{abstract}

\section{Introduction}

In {\em multi-agent path finding} (MAPF) \cite{DBLP:conf/focs/KornhauserMS84,DBLP:journals/jair/Ryan08,SharonSFS15,DBLP:journals/ai/SharonSGF13,DBLP:conf/aiide/Silver05,DBLP:conf/icra/Surynek09,DBLP:journals/jair/WangB11} the task is to navigate agents from given starting positions to given individual goals. The standard version of the problem takes place in undirected graph $G=(V,E)$ where agents from set $A=\{a_1,a_2,...,a_k\}$ are placed in vertices with at most one agent per vertex. The initial configuration of agents in vertices of the graph can be written as $\alpha_0: A \rightarrow V$ and similarly the goal configuration as $\alpha_+: A \rightarrow V$. The task of navigating agents can be expressed as a task of transforming the initial configuration of agents $\alpha_0: A \rightarrow V$ into the goal configuration $\alpha_+: A \rightarrow V$.

Movements of agents are instantaneous and are possible across edges into neighbor vertices assuming no other agent is entering the same target vertex. This formulation permits agents to enter vertices being simultaneously vacated by other agents. Trivial case when a pair of agents swaps their positions across an edge is forbidden in the standard formulation. We note that different versions of MAPF exist where for example agents always move into vacant vertices \cite{DBLP:journals/amai/Surynek17}. We usually denote the configuration of agents at discrete time step $t$ as $\alpha_t: A \rightarrow V$. Non-conflicting movements transform configuration $\alpha_t$ {\em instantaneously} into next configuration  $\alpha_{t+1}$. We do not consider what happens between $t$ and $t+1$ in this discrete abstraction. Multiple agents can move at a time hence the MAPF problem is inherently parallel.

In order to reflect various aspects of real-life applications variants of MAPF have been introduced such as those considering {\em kinematic constraints} \cite{DBLP:conf/ijcai/HonigK00XAK17}, {\em large agents} \cite{LargeAAAI2019}, or {\em deadlines} \cite{DBLP:conf/ijcai/0001WFLKK18} - see \cite{DBLP:journals/corr/0001KA0HKUXTS17} for more variants.

Particularly in this work we are dealing with an extension of MAPF that generalizes the constraint of having at most one agent per vertex. There are many situations where we need to model nodes that could hold more than agent at a time. Such situations include various graph-based evacuation models where for example nodes correspond to rooms in evacuated buildings \cite{DBLP:conf/aaai/KumarRH16} which naturally can hold more than one agent. Various spatial projections could also lead to having multiple agents per vertex such as upper projection of agents representing aerial drones where a single node corresponds to $x$,$y$-coordinate that could hold multiple agents at different $z$-coordinates \cite{DBLP:journals/corr/abs-1810-03071}. Generally the need to consider nodes capable of containing multiple agents appears in modeling of multi-agent motion planning task at higher levels of granularity.

\subsection{Contributions}

The contribution of this paper consists in showing how to generalize existing {\em propositional satisfiability} (SAT) \cite{Biere:2009:HSV:1550723} models of MAPF for finding optimal plans with general capacity constraints that bound the number of agents in vertices. Two existing SAT-based models are generalized: MDD-SAT \cite{SurynekFSB16} that builds the propositional model in an {\em eager way} and SMT-CBS \cite{DBLP:journals/corr/abs-1809-05959,Surynek_IJCAI-2019} that builds the model in a {\em lazy way} inspired by satisfiability modulo theories (SMT) \cite{DBLP:journals/jacm/NieuwenhuisOT06}.

The eager style of building the propositional model means that all constraints are posted into the model in advance. Such model is {\em complete} that is, it is solvable (satisfiable) if and only if the instance being modeled is solvable. In contrast to this, the lazy style does not add all constraints at once and works with {\em incomplete} models. The incomplete model preserve only one-sided implication w.r.t. solvability: if the instance being modeled is solvable then the incomplete model is solvable (satisfiable).

The SMT-CBS algorithm iteratively refines the incomplete model towards the complete one by eliminating conflicts. That is, a candidate solution is extracted from the satisfied incomplete model. The candidate is checked for conflicts - whether any of the MAPF rules is violated - for example if a collision between agents occurred. If there are no conflicts, we are finished as the candidate is a valid solution of the input MAPF instance. If a conflict is detected, then a constraint that eliminates this particular conflict is added to the incomplete model resulting in a new model and the process is repeated. That is, a new candidate solution is extracted from the new model etc. Eventually the process may end up with a complete model after eliminating all possible conflicts. However, we hope that the process finishes before constructing a complete model and we solve the instance with less effort.

In the presented generalization with capacity constraints we need to distinguish between the eager and lazy variant. The capacity constraint concerning given vertex $v$ bounding the number of agents that can simultaneously occupy $v$ by some integer constant say $2$ can be literally translated into the requirement that no $3$ distinct agents can occupy $v$ at the same time. Such a constraint can be directly posted in the eager variant: we either forbid all possible triples of agents in $v$ or post the corresponding {\em cardinality constraint} \cite{DBLP:conf/cp/BailleuxB03,DBLP:conf/cp/SilvaL07}.

The situation is different in the lazy variant. To preserve the nature of the lazy approach we cannot post the capacity bound entirely as conceptually at the low level as we are informed only about a particular MAPF rule violation, say for example agents $a_1$, $a_5$ and $a_8$ occurred simultaneously in $v$ which is forbidden in given MAPF. The information that there is a capacity constraint on $v$ bounding the number of agents in $v$ by $2$ may even not be accessible at the low level. Hence we can forbid simultaneous occurrence of only the given triple of agents, $a_1$, $a_5$ and $a_8$ in this case.

The paper is organized as follows. We first introduce the standard multi-agent path finding problem formally including commonly used objectives. Then we introduce two major existing SAT-based encodings. On top of this, we show how to extend these encodings with vertex capacities. Finally we evaluate extended models on standard benchmarks including open grids and large game maps.

\section{Formal Definition of MAPF and Vertex Capacities}

The {\em Multi-agent path finding} (MAPF) problem \cite{DBLP:conf/aiide/Silver05,DBLP:journals/jair/Ryan08} consists of an undirected graph $G=(V,E)$ and a set of agents $A=\{a_1, a_2, ..., a_k\}$ such that $|A| \leq |V|$. Each agent is placed in a vertex so that at most one agent resides in each vertex. The placement of agents is denoted $\alpha: A \rightarrow V$.  Next we are given initial configuration of agents $\alpha_0$ and goal configuration $\alpha_+$.

At each time step an agent can either {\em move} to an adjacent vertex or {\em wait} in its current vertex. The task is to find a sequence of move/wait actions for each agent $a_i$, moving it from $\alpha_0(a_i)$ to $\alpha_+(a_i)$ such that agents do not {\em conflict}, i.e., do not occupy the same location at the same time nor cross the same edge in opposite directions simultaneously. The following definition formalizes the commonly used movement rule in MAPF.

\begin{definition}
    {\bf Valid movement in MAPF.}
    Configuration $\alpha'$ results from $\alpha$ if and only if the following conditions hold:
    
    \begin{enumerate}[label=(\roman*)]
      \item {$\alpha(a) = \alpha'(a)$ or $\{\alpha(a),\alpha'(a)\} \in E$ for all $a \in A$ (agents wait or move along edges);}
      \item {for all $a \in A$ it holds ${\alpha(a) \neq \alpha'(a)} \Rightarrow {\neg (\exists b \in A)(\alpha(b) = \alpha'(a) \wedge \alpha'(b) = \alpha(a))}$ (no two agents crosses an edge in opposite directions);}
      \item {and for all $a,a' \in A$ it holds that ${a \neq a'} \Rightarrow {\alpha'(a) \neq \alpha'(a')}$ (no two agents share a vertex in the next configuration)}.
    \end{enumerate}
    \label{def:movement}
    \vspace{-0.4cm}
\end{definition}

Solving the MAPF instance is to find a sequence of configurations $[\alpha_0,\alpha_1, ...,\alpha_{\mu}]$ such that  $\alpha_{i+1}$ results using valid movements from $\alpha_{i}$ for $i=1,2,...,\mu-1$, and $\alpha_{\mu}=\alpha_+$.

A version of MAPF with {\em vertex capacities} generalizes the above definition by adding capacity function $c: V \rightarrow \mathbb{Z}_{+}$ that assigns each vertex a positive integer capacity. The interpretation is that a vertex $v$ can hold up to the specified number of agents $c(v)$ at any time-step.

The definition of the valid movement will change only in point (iii) where instead of permitting at most one agent per vertex we allow any number of agents not exceeding the capacity of the vertex:

\begin{definition}
    {\bf Vertex capacities in MAPF.}  
    \begin{enumerate}[label=(\roman*')]
      \setcounter{enumi}{2}
      \item {for all $v \in V$ it holds that $|{a \:|\: \alpha'(a) = v}| \leq c(v)$ (the number of agents in each vertex does not exceed the capacity in the next configuration)}.
    \end{enumerate}
    \label{def:movement-c}
    \vspace{-0.2cm}
\end{definition}

\begin{figure}[h]
    \centering
    \includegraphics[trim={3.2cm 21cm 3.7cm 2.5cm},clip,width=1.0\textwidth]{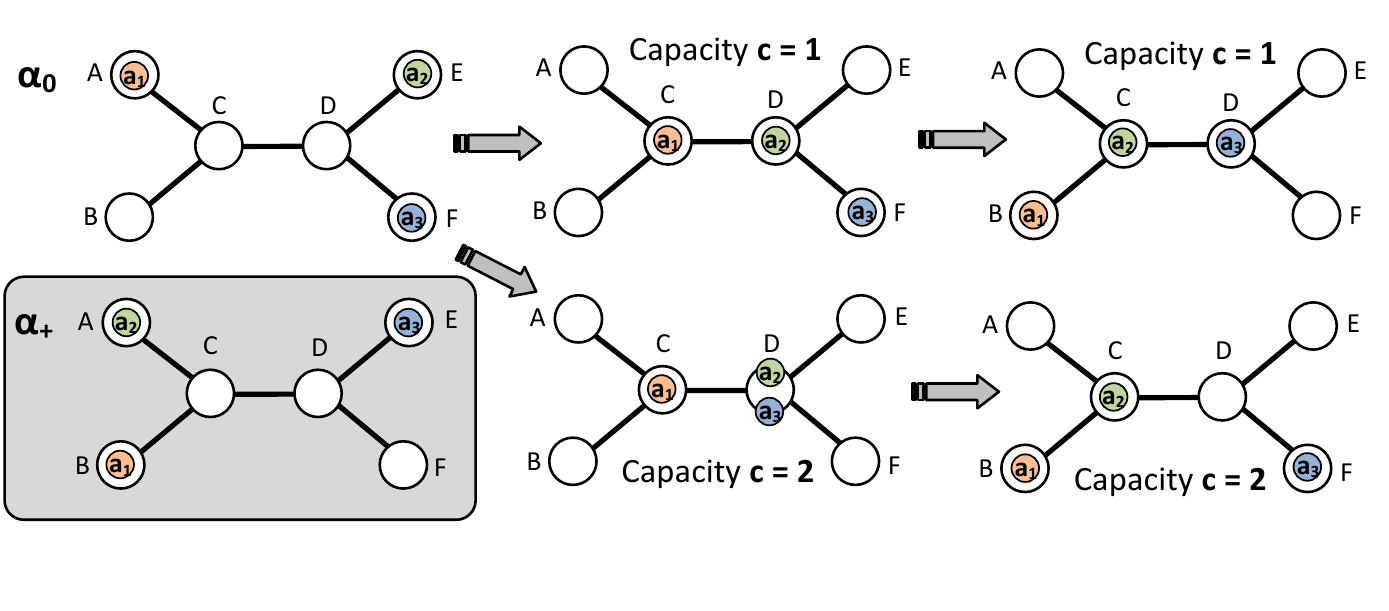}
    \vspace{-0.8cm}\caption{Illustration of the standard MAPF ($c=1$) and MAPF with generalized vertex capacity (uniform capacity $c=2$ us used). With $c=2$ two agents $a_2$ and $a_3$ can both enter vertex D. In contrast to this, $a_3$ must wait in vertex F in the standard MAPF.}
    \label{fig-MAPF-c}
\end{figure}

Using generalized vertex capacities relaxes the problem in fact as illustrated in Figure \ref{fig-MAPF-c}. Intuitively, capacities greater than one induce additional parking place in the environment which we hypothetise to make the problem easier to solve.

\subsection{Common Objectives in MAPF}

We address here optimal MAPF solving hence we need to introduce objective functions more formally. In case of {\em makespan} \cite{DBLP:journals/amai/Surynek17} we just need to minimize $\mu$ in the aforementioned solution sequence. For introducing the {\em sum-of-costs} objective \cite{dresner2008aMultiagent,standley2010finding,DBLP:journals/ai/SharonSGF13,SharonSFS15} we need the following notation:

\begin{definition}
	{\bf Sum-of-costs objective} is the summation, over all agents, of the number of time steps required to reach the goal vertex.
	Denoted $\xi$, where $\xi = \sum_{i=1}^k{\xi(path(a_i))}$, where $\xi(path(a_i))$ is an \textit{individual path cost} of agent $a_i$
	connecting $\alpha_0(a_i)$ calculated as the number of edge traversals and wait actions. \footnote{The notation $path(a_i)$ refers to
	path in the form of a sequence of vertices and edges connecting $\alpha_0(a_i)$ and $\alpha_+(a_i)$ while $\xi$ assigns the cost to a
	given path.}
\end{definition}

Observe that in the sum-of-costs we accumulate the cost of wait actions for agents not yet reaching their goal vertices. For the sake of brevity we focus here on the sum-of-costs, but we note that all new concepts can be introduced for different cumulative objectives like the makespan.\footnote{Dealing with objectives is out of scope of this paper. We refer the reader to \cite{SurynekFSB16} for more detailed discussion.}

We note that finding a solution that is optimal (minimal) with respect to the sum-of-costs objective is NP-hard \cite{DBLP:conf/aaai/RatnerW86}. The same result holds for the variant with capacities as it is a straight generalization of the standard MAPF version.

\section{Related Work}

Let us now recall existing SAT-based optimal MAPF solvers. We here focus on aspects important for introducing capacities. We recall MDD-SAT the sum-of-costs optimal solver based on {\em eager} SAT encoding \cite{SurynekFSB16} and SMT-CBS \cite{Surynek_IJCAI-2019}, the most recent SAT-based, or more precisely SMT-based, algorithm using lazy {\em encoding}.

\subsection{SAT-based Approach}

The idea behind the SAT-based approach is to construct a propositional formula $\mathcal{F(\xi)}$ such that it is satisfiable if and only if a solution of a given MAPF of sum-of-costs $\xi$ exists \cite{DBLP:journals/amai/Surynek17}. Moreover, the approach is constructive; that is, $\mathcal{F(\xi)}$ exactly reflects the MAPF instance and if satisfiable, solution of MAPF can be reconstructed from satisfying assignment of the formula. We say $\mathcal{F(\xi)}$ to be a {\em complete propositional model} of MAPF.

\begin{definition}
  {\bf (complete propositional model).} Propositional formula $\mathcal{F(\xi)}$ is a {\em complete propositional model} of MAPF $\Sigma$
  if the following condition holds:
  \begin{center}
  $\mathcal{F(\xi)}$ is satisfiable $\Leftrightarrow$ $\Sigma$ has a solution of sum-of-costs $\xi$.
  \end{center}
\end{definition}

Being able to construct such formula $\mathcal{F}$ one can obtain optimal MAPF solution by checking satisfiability of $\mathcal{F}(0)$, $\mathcal{F}(1)$, $\mathcal{F}(2)$,... until the first satisfiable $\mathcal{F(\xi)}$ is met. This is possible due to monotonicity of MAPF solvability with respect to increasing values of common cumulative objectives like the sum-of-costs. In practice it is however impractical to start at 0; lower bound estimation is used instead - sum of lengths of shortest paths can be used in the case of sum-of-costs. The framework of SAT-based solving is shown in pseudo-code in Algorithm \ref{alg-SAT}.

\subsection{Details of the MDD-SAT Encoding}

Construction of $\mathcal{F(\xi)}$ as used in the MDD-SAT solver relies on time expansion of underlying graph $G$. Having $\xi$, the basic variant of time expansion determines the maximum number of time steps $\mu$ ({\em makespan}) such that every possible solution of the given MAPF with the sum-of-costs less than or equal to $\xi$ fits within $\mu$ timesteps. Given $\xi$ we can calculate $\mu$ as $max_{i=1}^k\{\xi_0(a_i)\} + \xi - \xi_0$ where $\xi_0(a_1)$ is the length of the shortest path connecting $\alpha_0(a_i)$ and $\alpha_+(a_i)$; $\xi_0 = \sum_{i=1}^{k}{\xi_0(a_i)}$. The detailed justification of this equation is given in \cite{SurynekFSB16}.

%(that is, no agent is outside its goal vertex after $\mu$-th timestep and the sum-of-costs $\xi$ is not to be exceeded at the same time).

Time expansion itself makes copies of vertices $V$ for each timestep $t=0,1,2,...,\mu$. That is, we have vertices $v^t$ for each $v \in V$ and time step $t$. Edges from $G$ are converted to directed edges interconnecting timesteps in the time expansion. Directed edges $(u^t,v^{t+1})$ are introduced for $t=1,2,...,\mu-1$ whenever there is $\{u,v\} \in E$. Wait actions are modeled by introducing edges $(u^t,t^{t+1})$. A directed path in the time expansion corresponds to trajectory of an agent in time. Hence the modeling task now consists in construction of a formula in which satisfying assignments correspond to directed paths from $\alpha_0^0(a_i)$ to $\alpha_+^\mu(a_i)$ in the time expansion.

Assume that we have time expansion $TEG_i=(V_i,E_i)$ for agent $a_i$. Propositional variable $\mathcal{X}_v^t(a_j)$ is introduced for every vertex $v^t$ in $V_i$. The semantics of $\mathcal{X}_v^t(a_i)$ is that it is $\mathit{TRUE}$ if and only if agent $a_i$ resides in $v$ at time step $t$. Similarly we introduce $\mathcal{E}_u,v^t(a_i)$ for every directed edge $(u^t,v^{t+1})$ in $E_i$. Analogously the meaning of $\mathcal{E}_{u,v}^t(a_i)$ is that is $\mathit{TRUE}$ if and only if agent $a_i$ traverses edge $\{u,v\}$ between time steps $t$ and $t+1$.

Constraints are added so that truth assignment are restricted to those that correspond to valid solutions of a given MAPF. Added constraints together ensure that $\mathcal{F(\xi)}$ is a {\em complete propositional model} for given MAPF.

We here illustrate the model by showing few representative constraints. We omit here constraints that concern objective function. For the detailed list of constraints we again refer the reader to \cite{SurynekFSB16}.

Collisions among agents are eliminated by the following constraint for every $v \in V$ and timestep $t$ expressed on top of $\mathcal{X}_v^t(a_i)$ variables:

\begin{equation}
    {\sum_{a_i \in A \:|\:v^t \in V_i}{\mathcal{X}^t_v(a_i)} \leq 1
    }
    \label{eq-1}
\end{equation}

There are various ways how to translate the constraint using propositional clauses. One efficient way is to introduce $\neg \mathcal{X}^t_v(a_i) \vee \neg \mathcal{X}^t_v(a_j)$ for all possible pairs of $a_i$ and $a_j$.

Next, there is a constraint stating that if agent $a_i$ appears in vertex $u$ at time step $t$ then it has to leave through exactly one edge $(u^t,v^{t+1})$. This can be established by following constraints:

\begin{equation}
   {  \mathcal{X}_u^t(a_i) \Rightarrow \bigvee_{(u^t,v^{t+1}) \in E_i}{\mathcal{E}^t_{u,v}(a_i),}
   }
   \label{eq:basic-start}
\end{equation}
\begin{equation}
   {  \sum_{v^{t+1}\:|\:(u^t,v^{t+1}) \in E_i }{\mathcal{E}_{u,v}^t{(a_i)} \leq 1}
   }
   \label{eq-2}
\end{equation}

Similarly, the target vertex of any movement except wait action must be
empty. This is ensured by the following constraint for every $(u^t,v^{t+1}) \in E_i$:
\begin{equation}
   {  \mathcal{E}_{u,v}^t(a_i) \Rightarrow \bigwedge_{a_j \in A \:|\: a_j \neq a_i \wedge v^{t} \in V_j}{\neg \mathcal{X}_v^{t}(a_j)}
   }
   \label{eq-3}
\end{equation}

Other constraints ensure that truth assignments to variables per individual agents form paths. That is if agent $a_i$ enters an edge it must leave the edge at the next time step.

\begin{equation}
   {  \mathcal{E}^t_{u,v}(a_i) \Rightarrow \mathcal{X}^t_v(a_i) \wedge \mathcal{X}^{t+1}_v(a_i)
   }
   \label{eq-4}
\end{equation}

A common measure how to reduce the number of decision variables derived from the time expansion is the use of {\em multi-value decision diagrams} (MDDs) \cite{DBLP:journals/ai/SharonSGF13}. The basic observation that holds for MAPF is that an agent can reach vertices in the distance $d$ (distance of a vertex is measured as the length of the shortest path) from the current position of the agent no earlier than in the $d$-th time step. Analogical observation can be made with respect to the distance from the goal position.

Above observations can be utilized when making the time expansion of $G$. For a given agent, we do not need to consider all vertices at time step $t$ but only those that are reachable in $t$ timesteps from the initial position and that ensure that the goal can be reached in the remaining $\mu - t$ timesteps.

\begin{algorithm}[t]
\begin{footnotesize}
\SetKwBlock{NRICL}{SAT-Based ($G=(V,E),A,\alpha_0,\alpha_+)$}{end} \NRICL{
    $paths \gets$ $\{$shortest path from $\alpha_0(a_i)$ to $\alpha_+(a_i) | i = 1,2,...,k\}$ \\
    $\xi \gets \sum_{i=1}^k{\xi(N.paths(a_i))}$ \\
    \While {$\mathit{TRUE}$}{
        $\mathcal{F}(\xi) \gets$ encode$(\xi,G,A,\alpha_0, \alpha_+)$\\
        $assignment \gets$ consult-SAT-Solver$(\mathcal{F}(\xi))$\\
        \If {$assignment \neq$ UNSAT}{
        	$paths \gets$ extract-Solution$(assignment)$\\
        	\Return $paths$\\
        }
        $\xi \gets \xi + 1$\\
    }
} \caption{Framework of SAT-based MAPF solving} \label{alg-SAT}
\end{footnotesize}
\end{algorithm}

\subsection{Resolving Conflicts Lazily in SMT-CBS}

SMT-CBS is inspired by search-based algorithm CBS \cite{DBLP:conf/aaai/SharonSFS12,CBSJUR} that uses the idea of resolving conflicts lazily; that is, a solution of MAPF instance is not searched against the complete set of movement constraints that forbids collisions between agents but with respect to initially empty set of collision forbidding constraints that gradually grows as new conflicts appear. SMT-CBS follows the high-level framework of CBS but rephrases the process into propositional satisfiability in a similar way as done in formula satisfiability testing in {\em satisfiability modulo theory} paradigm \cite{DBLP:journals/jacm/NieuwenhuisOT06,DBLP:conf/cp/Nieuwenhuis10,DBLP:journals/constraints/BofillPSV12}.

The high-level of CBS searches a {\em constraint tree} (CT) using a priority queue in breadth first manner. CT is a binary tree where each node $N$ contains a set of collision avoidance constraints $N.\mathit{constraints}$ - a set of triples $(a_i,v,t)$ forbidding occurrence of agent $a_i$ in vertex $v$ at time step $t$, a solution $N.paths$ - a set of $k$ paths for individual agents, and the total cost $N.\xi$ of the current solution.

The low-level process in CBS associated with node $N$ searches paths for individual agents with respect to set of constraints $N.\mathit{constraints}$. For a given agent $a_i$, this is a standard single source shortest path search from $\alpha_0(a_i)$ to $\alpha_+(a_i)$ that avoids a set of vertices $\{v \in V|(a_i,v,t) \in N. \mathit{constraints}\}$ whenever working at time step $t$. For details see \cite{SharonSFS15}.

CBS stores nodes of CT into priority queue $\textsc{Open}$ sorted according to the ascending costs of solutions. At each step CBS takes node $N$ with the lowest cost from $\textsc{Open}$ and checks if $N.\mathit{paths}$ represent paths that are valid with respect to MAPF movements rules - that is, $N.\mathit{paths}$ are checked for collisions. If there is no collision, the algorithms returns valid MAPF solution $N.\mathit{paths}$. Otherwise the search branches by creating a new pair of nodes in CT - successors of $N$. Assume that a collision occurred between agents $a_i$ and $a_j$ in vertex $v$ at time step $t$. This collision can be avoided if either agent $a_i$ or agent $a_j$ does not reside in $v$ at timestep $t$. These two options correspond to new successor nodes of $N$ - $N_1$ and $N_2$ that inherit the set of conflicts from $N$ as follows: $N_1.\mathit{conflicts} = N.\mathit{conflicts} \cup \{(a_i,v,t)\}$ and $N_2.\mathit{conflicts} = N.\mathit{conflicts} \cup \{(a_j,v,t)\}$. $N_1.\mathit{paths}$ and $N_1.\mathit{paths}$ inherit paths from $N.\mathit{paths}$ except those for agents $a_i$ and $a_j$ respectively. Paths for $a_i$ and $a_j$ are recalculated with respect to extended sets of conflicts $N_1.\mathit{conflicts}$ and $N_2.\mathit{conflicts}$ respectively and new costs for both agents $N_1.\xi$ and $N_2.\xi$ are determined. After this, $N_1$ and $N_2$ are inserted into the priority queue $\textsc{Open}$.

\begin{algorithm}[h]
\begin{footnotesize}
\SetKwBlock{NRICL}{SMT-CBS ($\Sigma = (G=(V,E),A,\alpha_0,\alpha_+))$}{end} \NRICL{
    $conflicts \gets \emptyset$\\
    $paths \gets$ $\{path^*(a_i)$ a shortest path from $\alpha_0(a_i)$ to $\alpha_+(a_i) | i = 1,2,...,k\}$ \\
    $\xi \gets \sum_{i=1}^k{\xi(paths(a_i))}$ \\
    \While {$\mathit{TRUE}$}{
         $(paths,conflicts) \gets$ SMT-CBS-Fixed($conflicts,\xi,\Sigma$)\\
        \If {$paths \neq$ UNSAT}{
        	\Return $paths$\\
        }
        $\xi \gets \xi + 1$\\
    }
}   
 
\SetKwBlock{NRICL}{SMT-CBS-Fixed($conflicts,\xi,\Sigma$)}{end} \NRICL{
	    $\mathcal{H}(\xi) \gets$ encode-Basic$(conflicts,\xi,\Sigma)$\\
	    \While {$\mathit{TRUE}$}{
	        $assignment \gets$ consult-SAT-Solver$(\mathcal{H}(\xi))$\\
	        \If {$assignment \neq UNSAT$}{
	            $paths \gets$ extract-Solution$(assignment)$\\
	            $collisions \gets$ validate($paths$)\\
                   \If {$collisions = \emptyset$}{
                      \Return $(paths,conflicts)$\\
                   }
                   \For{each $(a_i,a_j,v,t) \in collisions$}{
                      $\mathcal{H}(\xi) \gets \mathcal{H}(\xi) \cup \{\neg \mathcal{X}_v^t(a_i) \vee \neg \mathcal{X}_v^t(a_j)$\}\\
                      $conflicts \gets conflicts \cup \{[(a_i,v,t),(a_j,v,t)]\}$
                   }
               }
               \Return {(UNSAT,$conflicts$)}\\
          }
}
\caption{SMT-CBS algorithm for solving MAPF} \label{alg-SMTCBS}
\end{footnotesize}
\end{algorithm}

SMT-CBS compresses CT into a single branch in which the propositional model taken from MDD-SAT is iteratively refined. The high-level branching from CBS is deferred to the low level of SAT solving. In the MDD-SAT encoding collision avoidance constraints are omitted initially, only constraints ensuring that assignments form valid paths interconnecting starting positions with goals are be preserved. This will result in an {\em incomplete propositional model} denoted $\mathcal{H}(\xi)$. The important component of SMT-CBS is a paths validation procedure that reports back the set of conflicts found in the current solution that are used for making model refinements. SMT-CBS is shown in pseudo-code as Algorithm \ref{alg-SMTCBS}.

The algorithm is divided into two procedures: SMT-CBS representing the main loop and SMT-CBS-Fixed solving the input MAPF for fixed cost $\xi$. The major difference from the standard CBS is that there is no branching at the high-level. The high-level SMT-CBS roughly correspond to the main loop of MDD-SAT. The set of conflicts is iteratively collected during the entire execution of the algorithm. Procedure {\em encode} from MDD-SAT is replaced with {\em encode-Basic} that produces encoding that ignores specific movement rules (collisions between agents) but in contrast to {\em encode} it encodes collected conflicts into $\mathcal{H}(\xi)$.

The conflict resolution in the standard CBS implemented as high-level branching is here represented by refinement of $\mathcal{H}(\xi)$ with disjunction (line 20). The presented SMT-CBS can eventually build the same formula as MDD-SAT but this is done lazily in SMT-CBS.

\section{Handling Capacity Constraints in MAPF}

To adapt the SAT-based approach for MAPF with capacities we need minor modifications only in both MDD-SAT and SMT-CBS. However in each algorithm the integration of capacity constraints in profoundly different. While in MDD-SAT we integrate capacity constraints eagerly in the line with the original design of the algorithm (that is, the constraint in introduced as a whole), in SMT-CBS we integrate capacity constraint lazily which means part by part as new conflicts appear.

\subsection{Details of the Encoding with Capacities}

We need only a small modification of the MDD-SAT encoding to handle vertex capacities. We need to replace constraint \ref{eq-1} with the following constraint that is again posted for every vertex $v$ and time step $t$:

\begin{equation}
    {\sum_{a_i \in A \:|\:v^t \in V_i}{\mathcal{X}^t_v(a_i)} \leq c(v)
    }
    \label{eq-10}
\end{equation}

Unlike in the standard MAPF we need here a more sophisticated translation of the constraint to propositional clauses. Using the approach of forbidding individual $c(v) + 1$-tuples can be highly inefficient especially in cases when $c(v)$ is large. Therefore we use cardinality constraints encodings commonly used in SAT \cite{DBLP:conf/cp/BailleuxB03,DBLP:conf/cp/Sinz05,DBLP:conf/cp/SilvaL07}. Generally the cardinality constraint over set of propositional variables $\{\mathcal{X}_1,\mathcal{X}_2,...,\mathcal{X}_n\}$ permits at most a specified number of variables from the set to be $\mathit{TRUE}$, denoted ${\leq}_k\{\mathcal{X}_1,\mathcal{X}_2,...,\mathcal{X}_n\}$ means that at most k variables from the set can be $\mathit{TRUE}$.

In our case of MAPF with capacities we need to introduce following cardinality constraints for every vertex $v$ and time step $t$. The practical implementation of cardinaliy constraints is done through encoding adder circuits inside the formula \cite{DBLP:conf/cp/SilvaL07}.

\begin{equation}
    {{\leq}_{c(v)}\{\mathcal{X}^t_v(a_i) \:|\: a_i \in A \wedge v^t \in V_i\}
    }
    \label{eq-11}
\end{equation}

\subsection{Capacities in SMT-CBS}

Capacities in SMT-CBS are resolved lazily as well. That is, the capacity constraint is not posted entirely as a cardinality constraint but instead individual sets of agents that violate the capacity are forbidden one by one as they appear. That is for example if a generalized conflict occurs with agents $a_{i_1}$, $a_{i_2}$, ..., $a_{i_m}$ in vertex $v$ (in other words if $m > c(v)$) we post a conflict elimination clause concerning the colliding set of agents: $\neg \mathcal{X}_v^t(a_{i_1}) \vee \neg \mathcal{X}_v^t(a_{i_2}) \vee ... \vee \neg \mathcal{X}_v^t(a_{i_m})$.

Hence in the SMT-CBS algorithm we modify only lines 20 and 21 that handle generalized vertex conflicts. Also we need to modify the validate procedure called at line 15 to reflect generalized vertex capacities.

\begin{figure}[t]
    \centering
    \includegraphics[trim={2.5cm 14.5cm 2cm 2.5cm},clip,width=0.95\textwidth]{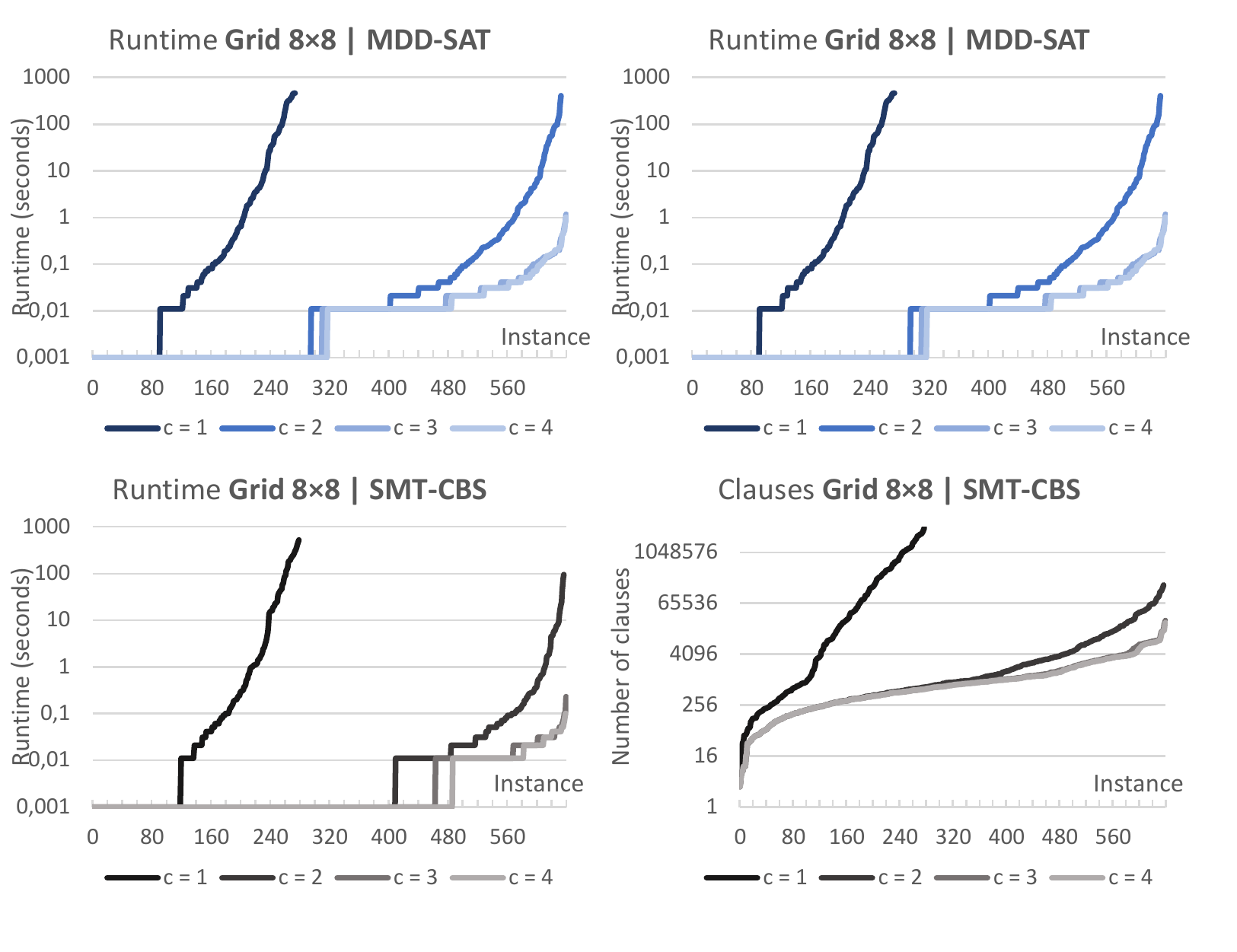}
    \vspace{-0.6cm}\caption{Sorted runtimes and the number of clauses on the $8 \times 8$ grid. MDD-SAT and SMT-CBS are compared.}
    \label{expr-grid-8x8}
\end{figure}

\section{Experimental Evaluation}

To evaluate the performance of capacity handling in context of SAT-based algorithms we performed an extensive evaluation on both standard synthetic benchmarks \cite{DBLP:conf/ijcai/BoyarskiFSSTBS15,DBLP:journals/ai/SharonSGF13} and large maps from games \cite{sturtevant2012benchmarks}.

\subsection{Setup of Experiments and Benchmarks}

We took the existing implementations of MDD-SAT and SMT-CBS written in C++. Both implementations are built on top of the Glucose 4 SAT solver \cite{DBLP:conf/sat/AudemardLS13,DBLP:conf/ijcai/AudemardS09}. In the implementations we modified the capacity constraint from the original {\em at-most-one} to generalized variants as mentioned above. All experiments were run on a Ryzen 7 CPU 3.0 Ghz under Kubuntu linux 16 with 16GB RAM. The timeout in all experiments was set to 500 seconds. Presented are only results finished within this time limit.

\begin{figure}[t]
    \centering
    \includegraphics[trim={2.5cm 14.5cm 2cm 2.5cm},clip,width=0.95\textwidth]{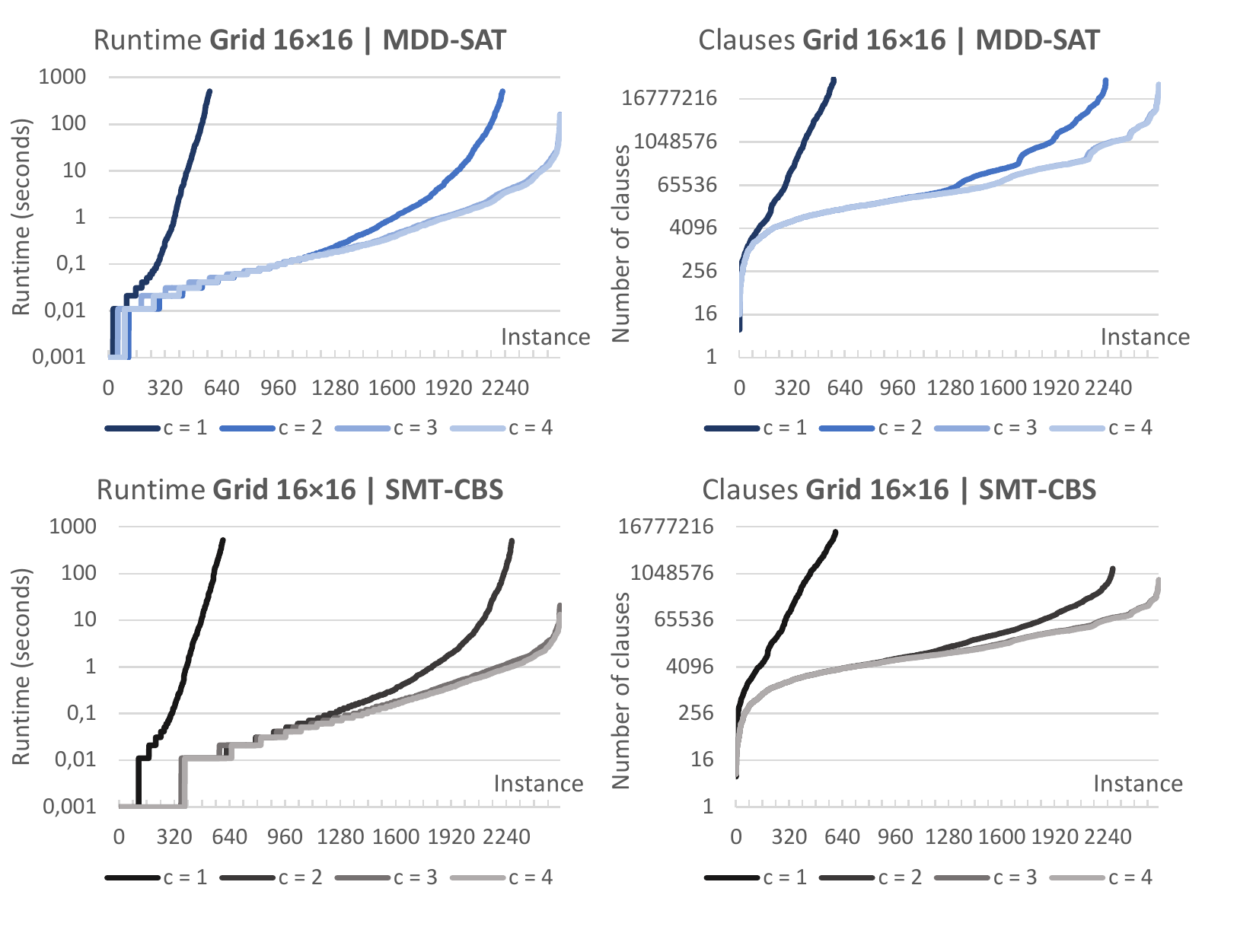}
    \vspace{-0.6cm}\caption{Sorted runtimes and the number of clauses on the $16 \times 16$ grid. MDD-SAT and SMT-CBS are compared.}
    \label{expr-grid-16x16}
\end{figure}

%\footnote{To enable reproducibility of results presented in this paper we provide complete source codes and experimental data on author's web page: \href{http://users.fit.cvut.cz/surynpav/research/aiia2019}{http://users.fit.cvut.cz/surynpav/research/aiia2019}.}.

The second part of experimental evaluation took place on large 4-connected maps taken from {\em Dragon Age} \cite{SharonSFS15,sturtevant2012benchmarks}. We took three structurally different maps focusing on various aspects such as narrow corridors, large almost isolated rooms, or topologically complex open space. In contrast to small instances, these were only sparsely populated with agents. Initial and goal configuration were generated at random again. Up to 80 agents were used in these instances and uniform capacities of 1, 2, and 3. On large maps we measured the runtime.

\subsection{Results on Small Grids}

Results obtained for small open grids are presented in Figures \ref{expr-grid-8x8} and \ref{expr-grid-16x16}. We can see that in comparison with the standard MAPF capacities bring significant reduction of difficulty of instances. This difference can be seen in both MDD-SAT and SMT-CBS. The starkest performance difference is between $c=1$ and $c=2$. The least performance difference is between $c = 3$ and $c=4$. The similar picture can be seen in for the number of clauses.

\begin{figure}[t]
    \centering
    \includegraphics[trim={1.5cm 21cm 1.5cm 2.5cm},clip,width=0.95\textwidth]{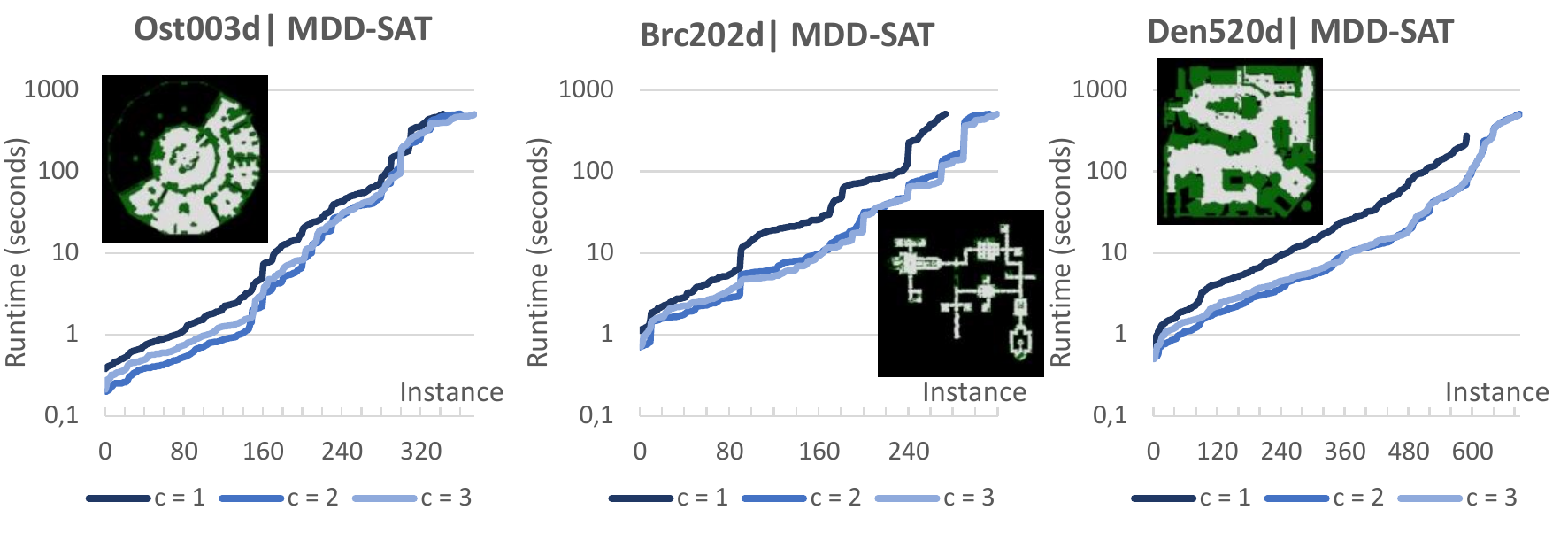}
    \vspace{-0.4cm}\caption{Sorted runtimes of MDD-SAT on \texttt{ost003d}, \texttt{brc202d}, and \texttt{den520d} maps.}
    \label{expr-maps-mddsat}
\end{figure}

\subsection{Results on Large Maps}

\begin{figure}[t]
    \centering
    \includegraphics[trim={1.5cm 21cm 1.5cm 2.5cm},clip,width=0.95\textwidth]{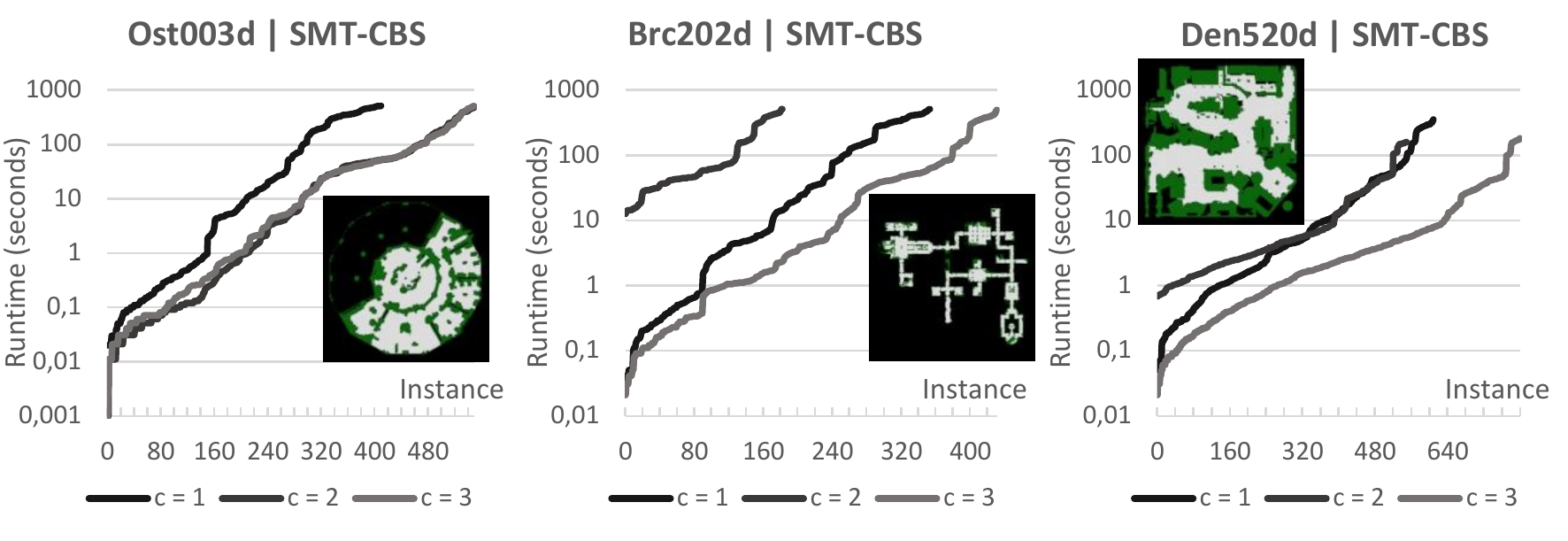}
    \vspace{-0.4cm}\caption{Sorted runtimes of MDD-SAT on \texttt{ost003d}, \texttt{brc202d}, and \texttt{den520d} maps.}
    \label{expr-maps-smtcbs}
\end{figure}

Results for large game maps are shown in Figures  \ref{expr-maps-mddsat} and \ref{expr-maps-smtcbs}. A different picture can be seen here. Adding capacities does not cause any significant simplification except the \texttt{brc202d} map which consists of narrow corridors. The interpretation is that adding extra parking place through capacities may lead to simplification only when it is not available normally. Otherwise generalized capacity constraints lead to harder instances.

\section{Discussion and Conclusion}

We introduced multi-agent path finding problem with vertex capacity constraints. We modified two existing state-of-the-art SAT-based optimal MAPF solvers to reflect vertex capacities, the MDD-SAT solver using the {\em eager} encoding and the SMT-CBS solver using the {\em lazy} encoding.

In both solvers we observed that adding an extra room by increasing the capacity of vertices dramatically reduces the difficulty of instances. However adding further capacity does has less significant effect. In large maps using higher capacities even lead to performance degradation which we attribute to more complex constraints.

In the future work we would like to apply the MAPF formulation with capacities in the real-life multi-robot problem.

\bibliographystyle{splncs04}
\bibliography{bibfile}

\end{document}